\documentstyle[12pt]{article}
\def\th{\theta}

\def\ka{\kappa}

\def\si{\sigma}

\def\ph{\phi}

\def\ps{\psi}

\def\Ps{\Psi}

\def\fr#1#2{{{#1} \over {#2}}}

\def\vev#1{\langle {#1}\rangle}

\def\half{{\textstyle{1\over 2}}}
\def\frac#1#2{{\textstyle{{#1}\over {#2}}}}

\def\lsim{\mathrel{\rlap{\lower4pt\hbox{\hskip1pt$\sim$}}
    \raise1pt\hbox{$<$}}}
\def\gsim{\mathrel{\rlap{\lower4pt\hbox{\hskip1pt$\sim$}}
    \raise1pt\hbox{$>$}}}
\def\sqr#1#2{{\vcenter{\vbox{\hrule height.#2pt
         \hbox{\vrule width.#2pt height#1pt \kern#1pt
         \vrule width.#2pt}
         \hrule height.#2pt}}}}

\newcommand{\beq}{\begin{equation}}
\newcommand{\eeq}{\end{equation}}
\newcommand{\bea}{\begin{eqnarray}}
\newcommand{\eea}{\end{eqnarray}}
\newcommand{\rf}[1]{(\ref{#1})}

\renewenvironment{thebibliography}[1]
 { \rm
   \begin{list}{\arabic{enumi}.}
    {\usecounter{enumi} \setlength{\parsep}{0pt}
     \setlength{\itemsep}{3pt} \settowidth{\labelwidth}{#1.}
     \sloppy
    }}{\end{list}}

\begin{document}
\titlepage

\begin{flushright}
{COLBY-96-02\\}
{IUHET 327\\}
{January 1996\\}
\end{flushright}
\vglue 1cm
	    
\begin{center}
{{\bf REVIVAL STRUCTURE OF STARK WAVE PACKETS\\}
\vglue 1cm
{Robert Bluhm,$^a$ V. Alan Kosteleck\'y,$^b$ and
Bogdan Tudose$^b$\\} 
\bigskip
{\it $^a$Physics Department\\}
\medskip
{\it Colby College\\}
\medskip
{\it Waterville, ME 04901, U.S.A.\\}
\bigskip
{\it $^b$Physics Department\\}
\medskip
{\it Indiana University\\}
\medskip
{\it Bloomington, IN 47405, U.S.A.\\}

}
\vglue 0.8cm

\end{center}

{\rightskip=3pc\leftskip=3pc\noindent
The revival structure of Stark wave packets is considered.
These wave packets have energies depending on two quantum
numbers and are characterized by two sets of classical periods
and revival times.
The additional time scales
result in revival structures different from those
of free Rydberg wave packets.
We show that Stark wave packets can exhibit fractional revivals.
We also show that these wave packets exhibit particular 
features unique to the Stark effect.
For instance,
the wave functions can be separated into
distinct sums over even and odd values of the principal
quantum number.
These even and odd superpositions interfere in different ways,
which results in unexpected periodicities in the interferograms
of Stark wave packets.

}

\vfill
\newpage

\baselineskip=20pt

A free Rydberg wave packet,
i.e., one evolving in the absence of external fields,
initially follows the motion of a charged particle 
in a Coulomb field.
However,
after several cycles it collapses
and a cycle of full and fractional revivals and
superrevivals commences
\cite{ps,az,ap,nau2,sr,bkp}.
This behavior holds not only for hydrogenic wave packets 
but also for wave packets in alkali-metal atoms
\cite{sr2},
where the energies have quantum defects 
causing shifts in the classical period 
and in the revival and superrevival times.
Theoretical descriptions of these wave packets
as squeezed states are known 
\cite{squeezed}.

The properties of Rydberg wave packets 
are being investigated
in the presence of strong fields,
which can significantly alter the atomic dynamics.
Recent experiments have examined the behavior of wave packets
in external electric and magnetic fields
\cite{noordam,walther,broers,wals,lankhui,bucksbaum}.
However,
the question of whether fractional revivals can occur in
these more complicated systems has yet to be addressed.

In this paper
we examine the revival structure of Stark wave packets,
which evolve in the presence of a static electric field.
These wave packets have energies that depend on two quantum
numbers.
We prove below that under certain conditions Stark wave
packets can exhibit full and fractional revivals.
Moreover,
we show the existence of new wave-packet behavior that does
not occur for free wave packets and that is experimentally
accessible.

To create a Stark wave packet,
an atom is first placed in a static electric field 
that splits and shifts the energy levels.
A short laser pulse is then applied
in the presence of the electric field,
resulting in a coherent superposition of Stark levels.
For a hydrogen atom in a small electric field, 
the energies in atomic units are 
$E_{n k} = - (1/{2 n^2}) + 3nkF/2$,
where $n$ is the principal quantum number,
$k=n_1 - n_2$ with $n_1$ and $n_2$ 
being parabolic quantum numbers,
and $F$ is the magnitude of the electric-field strength.

Stark wave packets have been produced and studied experimentally.
The production of wave packets consisting of 
a superposition of $k$ states in one Stark manifold 
with a fixed value of $n$ is described in 
ref.\ \cite{noordam}.
The oscillation of these parabolic wave packets corresponds
to an oscillation of the eccentricity of the orbit.
The time period of these oscillations is given by
$T_{\rm cl}^{(k)} = {2 \pi}/{3nF}$.

The dynamics of Stark wave packets above the classical
field-ionization threshold $F_c$ was examined
in refs.\ \cite{broers,lankhui}.
Simultaneous quantum beats in both the radial motion
and angular motion were observed,
with the period of the radial oscillations given by
the classical keplerian period 
$T_{\rm cl}^{(n)} = 2 \pi {\bar n}^3$.
Here,
$\bar n$ corresponds to the central energy excited
by the short laser pulse.
Note that the experimental observation time can extend through the  
lifetimes of the states comprising the wave packet.

Stark wave packets with longer lifetimes can be
created by forming combinations of states 
below the classical field-ionization threshold.
Superpositions of $k$ states with $n = 23$--$25$
have recently been produced in cesium
\cite{bucksbaum}.
Although the Stark spectra for alkali-metal atoms show
strong avoided crossings,
experiments indicate the behavior 
of these Stark wave packets 
is similar to ones in hydrogen.
This is because on average the energy spacings
between states of an alkali-metal atom are similar 
to those in hydrogen,
and it is these spacings that determine the motion
of the wave packet.

We are interested in the revival structure of a Stark wave packet 
$\Ps (t)$ formed as a coherent superposition
of states $\ph_{n k}$ with energies $E_{n k}$
depending on two quantum numbers $n$ and $k$.
We write 
$\Psi (t) = \sum_{n, k} c_{n k}
\ph_{n k} \exp \left[ -i E_{n k} t \right]$.
The quantum number $k$ is
even or odd according to whether $n$ is odd or even.
The two-unit jump of adjacent values of $k$ for fixed $n$
requires special handling 
in the treatment of fractional revivals and results
in additional interference effects in the interferograms
of Stark wave packets.
We suppose that the superposition is weighted around central
values $\bar n$ and $\bar k$ of the two quantum numbers.
The weighting coefficients $c_{n k}$ are therefore taken
to be strongly peaked around a central value 
$E_{\bar n \bar k}$ of the energy.
The energy can then be expanded in a Taylor series
around $E_{\bar n \bar k}$.

For definiteness,
consider an expansion centered around the values
$n = \bar n$ and $k = \bar k = 0$,
and take the quantum number $m$ associated with the third
component of the angular momentum to be zero.
In this case,
we introduce 
\beq
T_{\rm cl}^{(n)} 
= \fr {2 \pi} {\left( \fr {\partial E} {\partial n} 
\right)_{\bar n,\bar k}}
= 2 \pi \bar n^3
\quad , \quad\quad\quad
T_{\rm cl}^{(k)} 
= \fr {2 \pi} {2\left( \fr {\partial E} {\partial k} 
\right)_{\bar n,\bar k}}
= \fr {2 \pi} {3 F \bar n}
\quad ,
\label{Tclnk}
\eeq
\beq
t_{\rm rev}^{(n)} = \fr {2 \pi} {\fr 1 2 \left( 
\fr {\partial^2 E} {\partial n^2} 
\right)_{\bar n,\bar k}}
= \fr {4 \pi} 3 \bar n^4
\quad , \quad\quad\quad
t_{\rm rev}^{(nk)} = \fr {2 \pi} {2\left( 
\fr {\partial^2 E} {\partial n \partial k} 
\right)_{\bar n,\bar k}}
= \fr {2 \pi} {3 F}
\quad .
\label{trevnk}
\eeq
There is no revival time $t_{\rm rev}^{(k)}$ associated
with the quantum number $k$ since
${\partial^2 E}/{\partial k^2} = 0$.
Note that the definitions for $T_{\rm cl}^{(k)}$ and
$t_{\rm rev}^{(nk)}$ contain factors of two that
compensate for the two-unit jumps of adjacent $k$ values.
Note also that the mixed-derivative term generates
a time scale $t_{\rm rev}^{(n k)}$,
which we call the cross-revival time.
Substituting these definitions into $\Psi(t)$
and keeping terms to second order yields the expression
\beq
\Psi (t) = \sum_{n, k} c_{n k}
\ph_{n k} \exp \left[ -2 \pi i
\left(  \fr {(n - \bar n) t} {T_{\rm cl}^{(n)}}
+ \fr {k t} {2 T_{\rm cl}^{(k)}}
+ \fr {(n - \bar n)^2 t} {t_{\rm rev}^{(n)}}
+ \fr {(n - \bar n) k t} {2 t_{\rm rev}^{(nk)}}
\right)
\right]
\quad .
\label{psiexpans}
\eeq

For small $t$,
the first two terms of the time-dependent phase in 
Eq.\ \rf{psiexpans} dominate.
They represent beating between the two classical periods
$T_{\rm cl}^{(n)}$ and $T_{\rm cl}^{(k)}$.
We call $T_{\rm cl}^{(n)}$ and $T_{\rm cl}^{(k)}$
commensurate if
$T_{\rm cl}^{(n)} = a T_{\rm cl}^{(k)}/b$,
where $a$ and $b$ are relatively prime integers.
If this relation holds,
the time evolution of $\Ps (t)$ on short time scales
exhibits a period
$T_{\rm cl} = b T_{\rm cl}^{(n)} = a T_{\rm cl}^{(k)}$.

For larger times,
the revival time scales
$t_{\rm rev}^{(n)}$ and
$t_{\rm rev}^{(nk)}$ 
become relevant and modulate the initial behavior,
causing the wave packet to spread and collapse.
Examining the second-order terms in the
time-dependent phase, 
we find that the wave packet undergoes full revivals
provided the revival times 
$t_{\rm rev}^{(n)}$ and $t_{\rm rev}^{(nk)}$
are commensurate and obey
$t_{\rm rev}^{(n)} = rt_{\rm rev}^{(nk)}/s$,
where $r$ and $s$ are relatively prime integers.
If this relation is satisfied,
then there exists a revival time 
$t_{\rm rev} = s t_{\rm rev}^{(n)} = r t_{\rm rev}^{(nk)}$
at which both second-order terms 
in the phase are integer multiples of $2 \pi$.
Near $t_{\rm rev}$,
the phase is again controlled by the first-order terms,
and the shape and motion of the wave packet resembles
that of the initial wave packet,
i.e., 
a full revival occurs.

The commensurability of the time scales depends on 
$\bar n$ and $F$.
Restricting $F$ to 
below the classical field-ionization threshold 
$F_c = 1/ {16 \bar n^4}$
places limits on the ratios $a/b$ and $r/s$.
We find
$a/b < 3/{16}$
and
$r/s < 1/8$.
By tuning $F$,
specific commensurabilities and different types of revival 
structure can be selected.

An expression for the absolute square
of the autocorrelation function
$\vert A(t) \vert^2 = \vert \vev{\Ps(0) \vert \Ps(t)} \vert^2$ 
can be obtained
using the form of $\Ps(t)$ in
\rf{psiexpans} 
and the definitions \rf{Tclnk} and \rf{trevnk}
of the time scales.
The periodicities of the autocorrelation function reflect
the commensurabilities of the time scales.

To illustrate some of the possibilities,
consider two examples.
Let one have $a/b = 2/13$,
corresponding to a periodicity
$T_{\rm cl} = 2 T_{\rm cl}^{(k)}$,
while the other has $a/b = 1/6$,
corresponding to $T_{\rm cl} = T_{\rm cl}^{(k)}$.
In the first case,
peaks in the autocorrelation function should appear 
every two cycles in the period $T_{\rm cl}^{(k)}$,
while in the second peaks should appear every cycle.
Since $a/b = 3 F \bar n^4$,
these two different types of commensurability
can be obtained using two values of the field strength 
in the ratio 12/13.

Behavior of this type has been seen experimentally 
in Stark wave packets of cesium with $\bar n \simeq 24$
\cite{bucksbaum}.
Two different commensurabilities 
were observed in measured interferograms,
one with peaks every other $T_{\rm cl}^{(k)}$ cycle,
and the other with peaks every cycle.
The ratio of the two measured field strengths agrees 
with the ratio 12/13.

These features can be displayed in plots of
$\vert A(t) \vert^2$ as a function of $t$ for two
different values of $F$ having the ratio 12/13.
With $\bar n = 24$,
this requires using $F \simeq 794.8$ volts/cm for
$a/b = 2/13$ and $F \simeq 861.0$ volts/cm for
$a/b = 1/6$.
The superposition of interest lies near
$\bar n = 24$ with $m = 0$ and $\bar k =0$,
so the sum in $\vert A(t) \vert^2$ can be restricted
to the three $n$ values $n=23$, $24$, $25$
with only the upper part of the $n=23$ manifold 
and the lower part of the $n=25$ manifold included.
Weighting the $n$ states equally corresponds
to the flat-top distribution used in
ref.\ \cite{bucksbaum}.
For the distribution in $k$,
we choose a broad gaussian centered on $\bar k = 0$
with $\si_k = 6$.
This distribution matches the shape of the weighting coefficients
obtained in a short-pulse laser excitation from a low-energy
p state to a superposition of Stark states with s character.
A similar distribution would also occur in a superposition
of $m=1$ states in an excitation from a low-energy s state
to a superposition of Stark states with p character.

Figure 1 displays $\vert A(t) \vert^2$ as function of $t$
for the case where $a/b = 2/13$.
As expected,
the periodicity equals $2 T_{\rm cl}^{(k)}$,
with odd multiples suppressed.
Figure 2 shows the plot for $a/b = 1/6$.
In this case,
there are peaks every cycle with period $T_{\rm cl}^{(k)}$.
In both figures,
there is an overall decrease in the size of the peaks
as the time increases.
This is caused by the revival times,
which destroy the initial periodic motion.
Our analysis here uses hydrogenic energies,
and effects of core scattering to the continuum as would
occur in a Stark wave packet for an alkali-metal atom
are ignored.
A more detailed theoretical treatment
incorporating higher-order Stark effects and quantum defects
could be performed analytically using the methods of
ref.\ \cite{sqdt},
but this lies outside the scope of the present work.

For fractional revivals to form in Stark wave packets,
the wave function $\Psi(t)$ in Eq.\ \rf{psiexpans}
must be expressible as a sum of distinct subsidiary wave functions.
However,
this can only occur 
at times $t=t_{\rm frac}$
that are simultaneously irreducible rational fractions
of the two revival time scales.
We define
\beq
t_{\rm frac} = \fr {p_1} {q_1} t_{\rm rev}^{(n)}
= \fr {p_{12}} {q_{12}} t_{\rm rev}^{(nk)}
\quad .
\label{tfrac}
\eeq
Here,
the pairs of integers $(p_1,q_1)$
and $(p_{12},q_{12})$ are relatively prime.

In the Appendix,
we outline a proof that subsidiary waves form 
at the times $t_{\rm frac}$
and discuss some additional effects due to the
parity of $k$ and the extra factors of two in the phase.
We also show that $\Psi(t)$ in
\rf{psiexpans} can be written as a sum
$\Ps_{\rm odd}(t) + \Ps_{\rm even}(t)$
consisting of separate sums over odd and even
values of $n$.

At the fractional revivals,
each of the wave functions $\Ps_{\rm odd}(t)$ and
$\Ps_{\rm even}(t)$ can be written as a sum of subsidiary
waves $\ps^{\rm odd}_{\rm cl}$ and $\ps^{\rm even}_{\rm cl}$,
respectively,
with arguments shifted relative to $t$ by certain
fractions of the corresponding periods.
At the times $t = t_{\rm frac}$,
the result is 
\bea
\Psi(t) = \sum_{s_1 = 0}^{l_1-1} \sum_{s_2 = 0}^{l_2-1}
a^{\rm (odd)}_{s_1 s_2} 
\ps^{\rm (odd)}_{\rm cl} (t + \fr {s_1} {l_1} T_{\rm cl}^{(n)},
t + \fr {s_2} {l_2} T_{\rm cl}^{(k)})
\qquad\qquad\qquad\qquad\cr
+ \, e^{-i \pi \left( \fr {p_{12} t_{\rm rev}^{(nk)}} 
{q_{12} T_{\rm cl}^{(k)}} \right)} \,
\sum_{s_1 = 0}^{l^\prime_1-1} \sum_{s_2 = 0}^{l^\prime_2-1}
a^{\rm (even)}_{s_1 s_2} 
\ps^{\rm (even)}_{\rm cl} 
(t + \fr {s_1} {l^\prime_1} T_{\rm cl}^{(n)},
t + \fr {s_2} {l^\prime_2} T_{\rm cl}^{(k)})
\quad .
\label{starkfracrev}
\eea
The functions 
$\ps^{\rm odd}_{\rm cl}$ and $\ps^{\rm even}_{\rm cl}$,
the coefficients 
$a^{\rm (odd)}_{s_1 s_2}$ and $a^{\rm (even)}_{s_1 s_2}$, 
and the integers $l_1$, $l_2$, $l^\prime_1$, and $l^\prime_2$
are defined in the Appendix.

The functions $\ps^{\rm odd}_{\rm cl}$ and
$\ps^{\rm even}_{\rm cl}$ are doubly periodic functions
with periods $T_{\rm cl}^{\rm (n)}$ and $T_{\rm cl}^{\rm (k)}$.
The evolution of the wave packet
exhibits the beating of these two classical periods.
At the fractional revivals,
the sums in Eq.\ \rf{starkfracrev} exhibit
periodicities that are fractions of the time scales
$T_{\rm cl}^{\rm (n)}$ and $T_{\rm cl}^{\rm (k)}$.
The behavior of the quantum number $k$
causes the functions $\ps^{\rm odd}_{\rm cl}$ and
$\ps^{\rm even}_{\rm cl}$ to obey 
\beq
\ps^{\rm (odd)}_{\rm cl}(t + \half T_{\rm cl}^{(n)},t)
= - \ps^{\rm (odd)}_{\rm cl}(t,t)
\quad , \qquad 
\ps^{\rm (even)}_{\rm cl}(t + \half T_{\rm cl}^{(n)},t)
= \ps^{\rm (even)}_{\rm cl}(t,t)
\quad .
\label{per2}
\eeq
This additional dependence on $T_{\rm cl}^{\rm (n)}/2$
causes the unconventional revival structure of Stark wave packets.

As an illustrative example,
consider the case $\bar n = 24$,
and set $t_{\rm rev}^{(n)}/t_{\rm rev}^{(nk)} = r/s = 1/12$
by tuning the electric-field strength to $F \simeq 645.8$
volts/cm.
For this example,
$t_{\rm rev} = t_{\rm rev}^{(nk)} = 12 t_{\rm rev}^{(n)}$.
Using the expressions in the Appendix,
for $t \approx t_{\rm rev}$ we find 
$\Ps(t) \approx \ps^{\rm (odd)}_{\rm cl}(t,t) +
\ps^{\rm (even)}_{\rm cl}(t,t)$.
These sums are in phase and combine as a single total
wave packet,
producing the full revival at $t_{\rm rev}$.

At $t = t_{\rm rev}/2$,
however,
we find a time phase between $\Ps_{\rm odd}(t)$ and
$\Ps_{\rm even}(t)$,
with the full wave function reducing to
\beq
\Ps(t) \approx \ps^{\rm (odd)}_{\rm cl}
(t,t + \frac 1 2 T_{\rm cl}^{(k)})
 + \ps^{\rm (even)}_{\rm cl}(t + \frac 1 4 T_{\rm cl}^{(n)},t)
\quad .
\label{undieci}
\eeq
We see that this fractional revival
consists of two subsidiary wave functions out of phase
with each other.

Figure 3 shows the absolute square of the autocorrelation
function as a function of time.
Here,
$t_{\rm rev} \simeq 403.4$ psec,
$T_{\rm cl}^{(n)} \simeq 2.1$ psec,
and $T_{\rm cl}^{(k)} \simeq 16.8$ psec.
Since $T_{\rm cl}^{(k)}$ is an integer 
multiple of $T_{\rm cl}^{(n)}$,
we expect peaks at times equal to multiples of
$T_{\rm cl}^{(k)}$.
These are apparent in Fig.\ 3.
The full revival is evident
and has the anticipated periodicity.
The fractional revival near $t = t_{\rm rev}/2$
has peaks corresponding to 
those from two wave packets half a classical
period $T_{\rm cl}^{(k)}$ out of phase,
in agreement with our predictions.

For the odd-$n$ superposition in the Stark wave packet,
additional interference occurs at the 
$t =t_{\rm rev}/2$ revival.
The additional interference is caused 
by the antiperiodic behavior of
$\ps^{\rm (odd)}_{\rm cl}$ 
and can be seen explicitly from 
the form of the autocorrelation function.
The subsidiary wave functions $\ps^{\rm (odd)}_{\rm cl}$
and $\ps^{\rm (even)}_{\rm cl}$ are orthogonal since they
consist of separate sums over odd and even $k$.
We can therefore calculate $A(t) = \vev{\Ps(0) \vert
\Ps(t)}$ using Eq.\ \rf{undieci} 
for times $t \approx t_{\rm rev}/2$,
disregarding the cross terms.
This gives
\beq
A(t) =
\vev{\ps^{\rm (odd)}_{\rm cl}(0,0) \vert
\ps^{\rm (odd)}_{\rm cl}(t,t + \frac 1 2 T_{\rm cl}^{(k)})}
+ \vev{\ps^{\rm (even)}_{\rm cl}(0,0) \vert
\ps^{\rm (even)}_{\rm cl}(t + \frac 1 4 T_{\rm cl}^{(n)},t)}
\quad .
\label{bigA}
\eeq
Suppose the functions $\ps^{\rm (odd)}_{\rm cl}$
and $\ps^{\rm (even)}_{\rm cl}$ are spatially localized.
Then,
at times that are multiples of $T_{\rm cl}^{(k)}/2$,
$A(t)$ reduces to the first term 
since the second term vanishes.
Conversely,
at multiples of $T_{\rm cl}^{(k)}$,
$A(t)$ reduces to the second term.
Since $\ps^{\rm (odd)}_{\rm cl}$ is antiperiodic 
in the first time argument with period 
$T_{\rm cl}^{(n)}/2 \simeq 1.05$ psec,
we expect nodes in the autocorrelation function
occurring with this periodicity
at times that are multiples of $T_{\rm cl}^{(k)}/2$.
However,
since $\ps^{\rm (even)}_{\rm cl}$ is periodic,
nodes need not appear in $A(t)$ at 
multiples of $T_{\rm cl}^{(k)}$.

Figure 4 shows an enlargement of the autocorrelation
function near the fractional revival at
$t_{\rm rev}/2$.
Alternate peaks have different interference patterns,
as expected.
The peaks at multiples of $T_{\rm cl}^{(k)} \simeq 16.8$ psec
are single with no interference.
These arise from $\ps^{\rm (even)}_{\rm cl}$ in Eq.\ \rf{bigA}.
The peaks at multiples of $T_{\rm cl}^{(k)}/2$
arise from the $\ps^{\rm (odd)}_{\rm cl}$ terms in $A(t)$.
Nodes in $A(t)$ with the periodicity 
$T_{\rm cl}^{(n)}/2 \simeq 1.05$ psec are apparent.

The experiment described in
ref.\ \cite{bucksbaum} 
observed Stark wave packets for delay times of about 150 psec,
after which the signal was lost due to dephasing.
To detect the fractional revival
at $t_{\rm rev}/2$ described in this paper,
the alignment of the interferometer would need to be maintained
for delay times of at least 200 psec.
Observation of the full revival predicted
would require a delay time of 400 psec.
Delay times greater than these have already been achieved
in studies of Rydberg wave packets in the absence of
external fields.
Note that our treatment 
has disregarded core scattering and fine structure. 
The importance of these effects could therefore be determined 
in part by comparison of experiments to our predictions
for the fractional and full revivals.

\vglue 0.4cm

We thank Charlie Conover for useful discussions.
One of us (R.B.) would like to thank Colby College
for a Science Division grant.
This work is supported in part by the National
Science Foundation under grant number PHY-9503756.

\vglue 0.6cm
{\bf\noindent APPENDIX}
\vglue 0.4cm

This Appendix proves that fractional revivals
occur in Stark wave packets.
First, 
we rewrite $\Ps(t)$ in Eq.\ \rf{psiexpans}
by shifting $(n - \bar n) \rightarrow n$ and
separating the series into odd and even sums over $n$.
We then let
$k \rightarrow 2k$ in the sum over odd $n$, 
and $k \rightarrow 2k+1$ in the sum over even $n$.
This gives
$\Ps(t) = \Ps_{\rm odd}(t) + \Ps_{\rm even}(t)$,
where
\bea
\Ps_{\rm odd}(t) &=& \sum_{n \, {\rm odd}} \sum_{k} 
c_{nk} \ph_{nk} \,
\exp \left( -2 \pi i
\left(  \fr {n t} {T_{\rm cl}^{(n)}}
+ \fr {k t} {T_{\rm cl}^{(k)}}
- \fr {n^2 t} {t_{\rm rev}^{(n)}}
+ \fr {n k t} {t_{\rm rev}^{(nk)}}
\right)\right)
\quad ,\cr
\Ps_{\rm even}(t)
&=& e^{-i \pi ( t/ T_{\rm cl}^{(k)})} \,
\sum_{n \, {\rm even}} \sum_{k} 
c_{nk} \ph_{nk} 
\cr
\qquad \qquad \qquad 
&&\times 
\exp \left( -2 \pi i
\left(  \fr {n t} {T_{\rm cl}^{(n)}}
+ \fr {k t} {T_{\rm cl}^{(k)}}
- \fr {n^2 t} {t_{\rm rev}^{(n)}}
+ \fr {n k t} {t_{\rm rev}^{(nk)}}
+ \fr {n t} {2 t_{\rm rev}^{(nk)}}
\right)\right)
\quad .
\label{bigmess}
\eea
The shifted sums are now over integer values of $k$,
and the expansion coefficients $c_{nk}$ and the wave functions
$\ph_{nk}$ have been appropriately redefined.
Note the appearance of the additional phase terms in 
$\Ps_{\rm even}(t)$.

Define the doubly periodic wave functions
\beq
\ps^{\rm (odd)}_{\rm cl}(t_1,t_2) = 
\sum_{n \, {\rm odd}} \sum_{k} 
c_{nk} \ph_{nk} \,
\exp \left( -2 \pi i
\left(  \fr {n t_1} {T_{\rm cl}^{(n)}}
+ \fr {k t_2} {T_{\rm cl}^{(k)}}
\right)\right)
\quad ,
\label{psiclodd}
\eeq
\beq
\ps^{\rm (even)}_{\rm cl}(t_1,t_2) = 
\sum_{n \, {\rm even}} \sum_{k} 
c_{nk} \ph_{nk} \,
\exp \left( -2 \pi i
\left(  \fr {n t_1} {T_{\rm cl}^{(n)}}
+ \fr {k t_2} {T_{\rm cl}^{(k)}}
\right)\right)
\quad .
\label{psicleven}
\eeq
Then,
consider the periodicity in $n$ and $k$ for the
higher-order terms in the time-dependent phases of 
$\Ps_{\rm odd}(t)$ and $\Ps_{\rm even}(t)$ at 
$t = t_{\rm frac}$.
These terms are 
\beq
\th^{\rm (odd)}_{nk} = \fr {p_1} {q_1} n^2
- \fr r s \fr {p_1} {q_1} n k 
\quad ,
\label{thetaodd}
\eeq
\beq
\th^{\rm (even)}_{nk} = \fr {p_1} {q_1} n^2
- \fr r s \fr {p_1} {q_1} n k 
- \fr r s \fr {p_1} {q_1} \fr 1 2 n 
\quad .
\label{thetaeven}
\eeq
Here,
$n$ is odd in Eq.\ \rf{thetaodd} 
and even in Eq.\ \rf{thetaeven}.

We seek the minimum periods $l_1$, $l_2$, $l^\prime_1$, 
and $l^\prime_2$ such that
$\th^{\rm (odd)}_{n + l_1,k} = \th^{\rm (odd)}_{nk}$,
$\th^{\rm (odd)}_{n,k + l_2} = \th^{\rm (odd)}_{nk}$,
$\th^{\rm (even)}_{n + l^\prime_1,k} = \th^{\rm (even)}_{nk}$,
and
$\th^{\rm (even)}_{n,k + l^\prime_2} = \th^{\rm (even)}_{nk}$.
These relations yield four conditions 
for the periods $l_1$, $l_2$, $l^\prime_1$, and $l^\prime_2$
in terms of $n$, $k$, and $t_{\rm frac}$.

Since the functions 
$\ps^{\rm (odd)}_{\rm cl}$ and $\ps^{\rm (even)}_{\rm cl}$
with $t$ shifted by appropriate fractions of $T_{\rm cl}^{(n)}$
and $T_{\rm cl}^{(k)}$
have the same periodicities in $n$ and $k$  as
$\th^{\rm (odd)}_{nk}$ and $\th^{\rm (even)}_{nk}$,
respectively,
we may use these functions as a basis for an expansion
of the wave functions $\Ps_{\rm odd}(t)$ and $\Ps_{\rm even}(t)$
at the times $t_{\rm frac}$.
The result is the expansion 
\rf{starkfracrev}.

The expansion coefficients $a^{\rm (odd)}_{s_1 s_2}$
and $a^{\rm (even)}_{s_1 s_2}$ are 
\beq
a^{\rm (odd)}_{s_1 s_2} = \fr 1 {l_1 l_2}
\sum_{\ka_1 = 0}^{l_1-1} \sum_{\ka_2 = 0}^{l_2-1}
\exp \left( 2 \pi i \th^{\rm (odd)}_{\ka_1 \ka_2} \right)
\exp \left( 2 \pi i \fr {s_1} {l_1} \ka_1 \right)
\exp \left( 2 \pi i \fr {s_2} {l_2} \ka_2 \right)
\quad ,
\label{assodd}
\eeq
\beq
a^{\rm (even)}_{s_1 s_2} = \fr 1 {l^\prime_1 l^\prime_2}
\sum_{\ka_1 = 0}^{l^\prime_1-1} \sum_{\ka_2 = 0}^{l^\prime_2-1}
\exp \left( 2 \pi i \th^{\rm (even)}_{\ka_1 \ka_2} \right)
\exp \left( 2 \pi i \fr {s_1} {l^\prime_1} \ka_1 \right)
\exp \left( 2 \pi i \fr {s_2} {l^\prime_2} \ka_2 \right)
\quad ,
\label{asseven}
\eeq
When these expressions are substituted into
Eq.\ \rf{starkfracrev} and the definitions 
\rf{psiclodd} and \rf{psicleven} are used,
Eq.\ \rf{starkfracrev} reduces to the form given in
Eq.\ \rf{bigmess}.

\vglue 0.6cm
{\bf\noindent REFERENCES}
\vglue 0.4cm

\vfill\eject

\baselineskip=16pt
{\bf\noindent FIGURE CAPTIONS}
\vglue 0.4cm

\begin{description}
 
\item[{\rm Fig.\ 1:}]
The autocorrelation versus time in picoseconds for
a Stark wave packet with $\bar n = 24$.
The electric-field strength is $F \simeq 794.8$
volts/cm,
corresponding to the ratio
$T_{\rm cl}^{\rm (n)}/T_{\rm cl}^{\rm (k)} \simeq 2/13$.

\item[{\rm Fig.\ 2:}]
The autocorrelation versus time in picoseconds for
a Stark wave packet with $\bar n = 24$.
Here, the electric-field strength is $F \simeq 861.0$
volts/cm,
corresponding to the ratio
$T_{\rm cl}^{\rm (n)}/T_{\rm cl}^{\rm (k)} \simeq 1/6$.

\item[{\rm Fig.\ 3:}]
The revival structure of Stark wave packets is
displayed in a plot of the autocorrelation function as
a function of the time in picoseconds.
The Stark wave packet has $\bar n = 24$.
The electric-field strength is $F \simeq 645.8$
volts/cm,
which sets
$t_{\rm rev}^{\rm (n)}/t_{\rm rev}^{\rm (nk)} \simeq 1/12$ and
$T_{\rm cl}^{\rm (n)}/T_{\rm cl}^{\rm (k)} \simeq 1/8$.

\item[{\rm Fig.\ 4:}]
An enlargement of Fig.\ 3 in the vicinity of the
fractional revival at $t_{\rm rev}/2$.

\end{description}

\vfill
\eject
\end{document}